\documentclass[aps,pre,reprint,superscriptaddress,showpacs]{revtex4-1}

\usepackage{graphicx}
\usepackage{dcolumn}
\usepackage{bm}

\begin{document}


\title{Montecarlo simulation study of acid coagulation in casein suspensions}
\author{Tonatiuh Sosme-S\'anchez}
\affiliation{Facultad de Ciencias Qu\'{\i}micas, Universidad Aut\'onoma de Chihuahua. 
Circuito Universitario 1 S/N, Chihuahua, Chih., M\'exico}
\affiliation{Instituto Tecnol\'ogico Superior de San Andr\'es Tuxtla. 
Carretera Costera del Golfo S/N, San Andr\'es Tuxtla, Ver., M\'exico}

\author{N\'estor Guti\'errez-M\'endez}
\author{Jos\'e Manuel N\'apoles-Duarte}
\author{Luz Mar\'{\i}a Rodr\'{\i}guez-Valdez}
\author{Marco Antonio Ch\'avez-Rojo}
\email{mchavezrojo@gmail.com}
\affiliation{Facultad de Ciencias Qu\'{\i}micas, Universidad Aut\'onoma de Chihuahua. 
Circuito Universitario 1 S/N, Chihuahua, Chih., M\'exico}

\date{\today}

\begin{abstract}
In this work, we present a computer simulation study of the acid coagulation of casein micelles. 
For different conditions, we predict the acidity (pH value) at which aggregation transition occurs. 
The interaction potential implemented consists of three contributions, 
the electrostatic repulsion, the steric repulsion between the polyelectrolyte brush formed by the 
$\kappa$-casein exposed segments, and finally, the van der Waals attraction which is responsible 
for aggregation when the repulsive terms are attenuated by the positive ions present in the system. 
In our simulations, we employed the radial distribution function as an indicator of the 
aggregation transition. 
The predictions of the model
are summarized in diagrams showing the dependence of the isoelectric point on temperature, 
micelles volume fraction and surface concentration of $\kappa$-casein. 
\end{abstract}

\pacs{82.20.Wt, 82.70.Dd, 82.70.Uv, 87.15.bk}

\maketitle

\section{Introduction}
From a physicochemical perspective, the most important process in the elaboration 
of dairy products is casein coagulation \cite{fox2015cheese}. 
Cow milk is composed mainly of water ($\sim$87\%), lactose ($\sim$5\%), 
fat ($\sim$4\%), ashes ($\sim$1\%) and proteins ($\sim$3\%). 
Among these proteins, the most significant owing to their proportion 
and commercial importance are caseins;
they represent 80$\%$ of the total of protein content of milk \cite{fox2003p}.
Because of that, caseins have been studied extensively and are very well 
characterized \cite{de2003casein}.
They are composed of four fractions, the so-called $\alpha_{s1}-$,
$\alpha_{s2}-$, $\beta-$ and $\kappa-$. Being the latter the only soluble fraction which
represents less than 15$\%$.

Since most of the casein fractions are insoluble, they are assembled into colloidal 
casein-calcium-transport complexes termed casein micelles \cite{fox2008micelles} (CM).
Because many physicochemical properties of milk
are determined by the microscopic properties of casein micelles, 
it is important to understand how the interactions between them determines its macroscopic properties. 
With the purpose of explaining the synthesis, internal structure and stability
of the casein micelle, some models have been developed in the last 50 years. 
The interested reader may follow the development of these models in some recent reviews 
\cite{phadungath2005casein,farrell2006casein,qi2007studies,ferrandini2008modelos} and
may found an excellent discussion about the characteristics that a consistent model should contain in 
Ref. [\onlinecite{horne2006casein}].

Even though there are crucial differences between them and some important discrepancies
in the interpretation of experimental results \cite{horne2003casein,ouanezar2012afm}, 
the disagreement is concerning to the internal structure and interactions at the nanoscopic domain. 
But, from a colloidal point of view, there is general agreement that skim milk can be 
considered as a suspension of barely spherical micelles with sizes ranging from
50 to 500 nm, and sterically stabilized
by a polymer brush made up mainly of $\kappa$-casein ($\kappa$-Cn) hidrophilic segments 
\cite{fox2008micelles, de1996kappa}. 

Casein stability is due to the competence between attractive van der Waals forces 
and repulsive interactions, specifically, electrostatic and steric forces. 
De Kruif has proposed \cite{de1996kappa,de1999casein} that the steric repulsion 
of $\kappa$-casein segments is the most important at normal milk conditions and it dominates, 
preventing aggregation. 
When repulsive forces are attenuated, coagulation occurs. 
This attenuation can be achieved by two ways. 
In dairy products technology, these mechanisms are called 
enzymatic and acid coagulation \cite{fox2015cheese}, respectively. 
In the first case, employed in the production of all types of cheese, 
the destabilization is promoted by the activity of proteolytic enzymes
that cleave off the exposed segments of $\kappa$-casein, 
which leads to an attenuation of the steric repulsion, and so, 
van der Waals attraction causes micelles to aggregate. 
In the second case, employed mainly in the production of 
yogurth \cite{lucey1997formation,lee2010formation}, 
destabilization is achieved by a decrease in pH.
In this case, the increase in proton concentration, 
screens negative charge of the $\kappa$-casein segments, 
which makes them more flexible and less repulsive.
If this acidification goes beyond certain critic value, 
corresponding to the $\kappa$-casein isoelectric point, 
van der Waals attraction will induce aggregation. 
The isoelectric point of caseins is located at pH=4.6, 
which is the value at which milk coagulates.

From the colloidal perspective, the Adhesive Hard Sphere (AHS) model has been successfully 
employed \cite{de1992casein,de1997skim} in the prediction of the 
{\em initial} \cite{de1998supra, de1999casein, horne2003casein}
stages of the aggregation process, both for rennet- or acid-induced coagulation. 
Going beyond in the description of the aggregation process would require
the developing of more elaborate models, capable of predicting the
kinetics of the process, particularly the development of viscoelastic properties 
as a function of time \cite{bauer1995structure, horne2003casein, zhong2007physicochemical}.
It seems difficult to develope a colloidal-scale model with this characteristics,
that is why the more promisory proposals rely on the nanoscopic 
domain \cite{horne2006casein}. 
Nevertheless, colloidal models as the mentioned AHS can still be useful in
predicting physicochemical conditions affecting micelle-micelle interactions.


In this work, we employed a colloidal model proposed by
De Kruif \cite{de1999casein} in order to predict the conditions 
at which aggregation of the casein micelles occurs
by performing Montecarlo simulations.
Even though the interaction potential proposed in Ref. [\onlinecite{de1999casein}] 
cannot tell anything about the internal reestructuring of the micelle clusters, 
it seems to describe correctly what happens when casein micelles collide with each 
other \cite{tuinier2002stability, de2005stabilization}.
Hence, preliminary results presented in this work are promisory because computer simulations
can serve as a usefull tool in the selection of optimal conditions for dairy products ellaboration.
   

This paper is organized as follows: 
In section 2 the model and computer simulation details are described. 
In section 3 we present the aggregation diagrams for 
different micelle volume fraction, temperature, and grafting density, 
in order to study how the concentration, the kinetic energy, 
and the surface concentration of $\kappa$-casein in micelles surface, 
would affect the acid coagulation of skim milk. 
Conclusions and perspectives are presented in section 4.

\section{Model System}
The model system consists of a polydisperse suspension of spherical particles 
with an interaction potential with three contributions: electrostatic, steric and van der Waals.  

\subsection{Size distribution}
As mentioned above, the diameter of casein micelles ranges from 50 to 500 nm. 
In order to include the effects of polydispersity 
in the simulations, we implemented the log-normal distribution
reported in Ref. [\onlinecite{de1998supra}], 
\begin{equation}
f(d)=\frac{1}{\sqrt{2\pi}\beta d}e^{-\frac{(ln(d/d_0))^2}{2\beta^2}},
\label{distribution}
\end{equation}
where $d_0=200$ is the mean diameter of CM and $\beta=0.38$ is the standard deviation
of the distribution.

\subsection{Interaction potential}
Tuinier and Kruif proposed a potential for modelling CM interactions in Ref. [\onlinecite{tuinier2002stability}].
This potential considers the more significant contributions to the interaction energy  
between CM, these are, the van der Waals attraction, $U_{vdw}$, 
the electrostatic repulsion due to the surface charge of the micelles, $U_{el}$ 
and the steric repulsion between the polyelectrolyte brush formed by the exposed segments of $\kappa$-Cn, $U_{st}$. 

The Van der Waals attraction between two spheres of radii $a_1$ and $a_2$ 
is given by \cite{hamaker1937london}
\begin{eqnarray}
\beta U_{vdw}=A\Biggl[&&\frac{2a_1a_2}{r^2-(a_1+a_2)^2}+
\frac{2a_1a_2}{r^2-(a_1-a_2)^2}+\nonumber\\
&& +ln\biggl(\frac{r^2-(a_1+a_2)^2}{r^2-(a_1-a_2)^2}\biggr)\Biggr],
\end{eqnarray}
where $r$ is the distance between the center of the particles, 
$a_i$ is the radius of particle $i$ and $A$ is the Hamaker constant. 
This attraction is always present and so, when repulsive interactions
disappear, it is responsible of aggregation.

The first of the repulsion terms is the electrostatic repulsion
which is obtained from the  solution of the linearized Poisson-Boltzmann 
equation for two charged spheres surrounded by its counterions \cite{derjaguin1940repulsive,
verwey1999theory},
\begin{equation}
\beta U_{el}=\epsilon_0 \epsilon \varphi_1 \varphi_2 
ln(1+e^{r-a_1-a_2}),
\end{equation}
where $\epsilon_0$ is the electric permittivity of vacuum, 
$\epsilon$ is the dielectric constant of the solvent, 
$\varphi_i$ is the electrostatic potential at the surface of 
particle $i$ and $\kappa$ is the inverse Debye length. 

The last contribution is the steric repulsion between the polyelectrolyte 
brushes formed by de exposed segments of $\kappa$-Cn, 
that for two spherical brushes are given by \cite{likos2000colloidal}
\begin{eqnarray}
\beta U_{st}=\frac{16\pi a_{eff}H}{35\sigma}\Biggl[&& 28(y^{-1/4}-1)+\frac{20}{11}(1-y^{11/4})+\nonumber\\
&&+12(y-1)\Biggr]
\end{eqnarray}
where $a_{eff}$ is the effective radius defined as 
$a_{eff}=[\frac{1}{a_1}+\frac{1}{a_2}]^{-1}$,  
$H$ is the width of the polyelectrolyte brush 
\cite{braithwaite1999study,israels1994theory,israels1994charged,alexander1977adsorption,alexander1977polymer},
$y=\frac{r-a_1-a_2}{2H}$ is the distance between the surface of each 
sphere in units of $H$, and $\sigma$ is the grafting density.

\begin{figure}
\includegraphics [width=8cm]{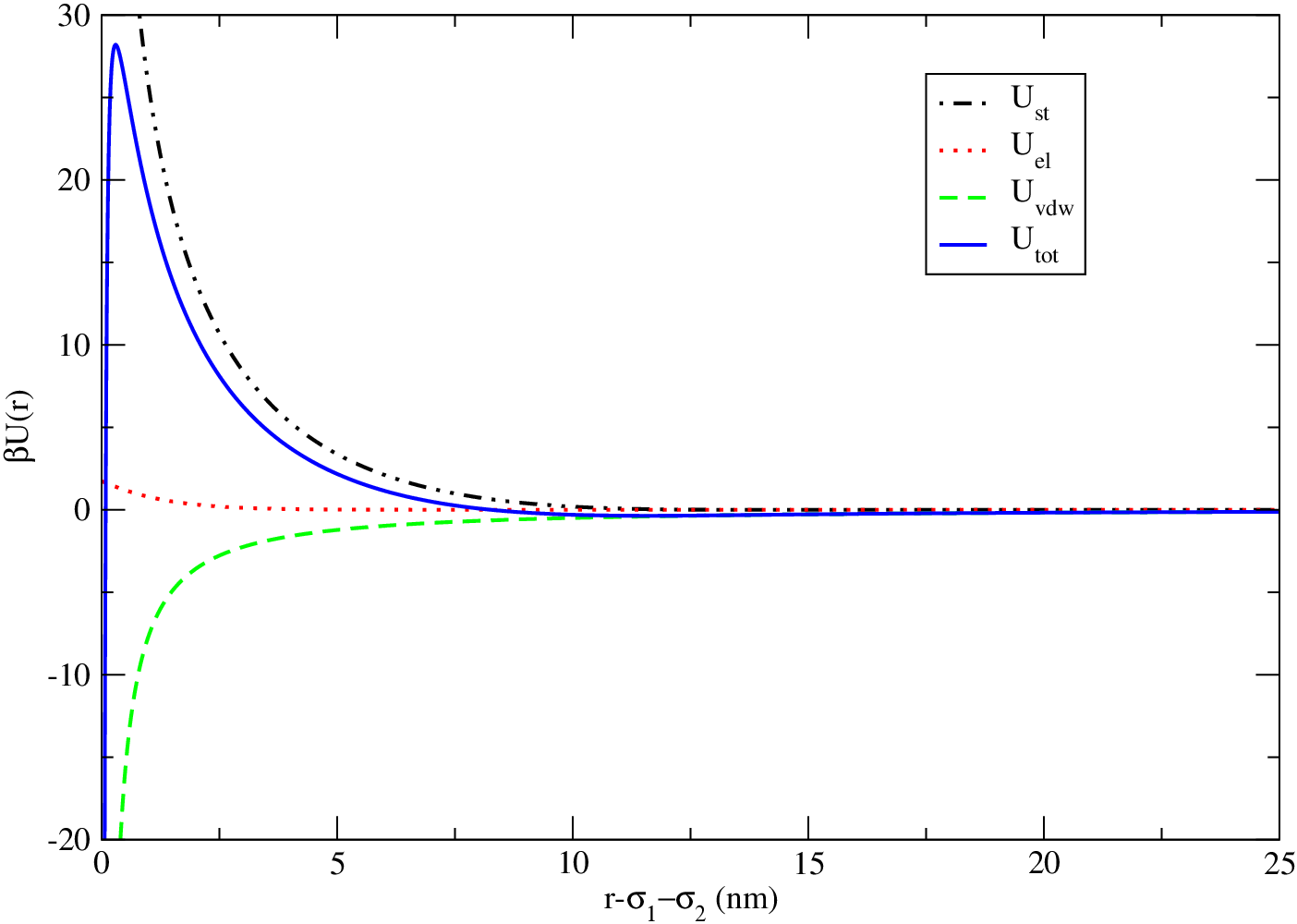}
\caption{Interaction potential between casein micelles with $r=100$ nm for $pH=6.7$. Steric (dot-dashed line), 
electrostatic (dotted line) and van der Waals (dashed line) contributions are plotted separately. 
The total interaction is plotted as a continuous line.}
\label{pot}
\end{figure}

These three terms are plotted in Fig.\ref{pot}  for two micelles with diameter 
$a_1=a_2=100 nm$, and for $pH=6.7$, which corresponds to the acidity of milk. 
At these conditions, it is shown that the steric repulsion is much greater 
than electrostatic repulsion and several times greater than the van der Waals attraction, 
preventing CM aggregation, which explains the stability of milk. 
 
\begin{figure}
\includegraphics [width=8cm]{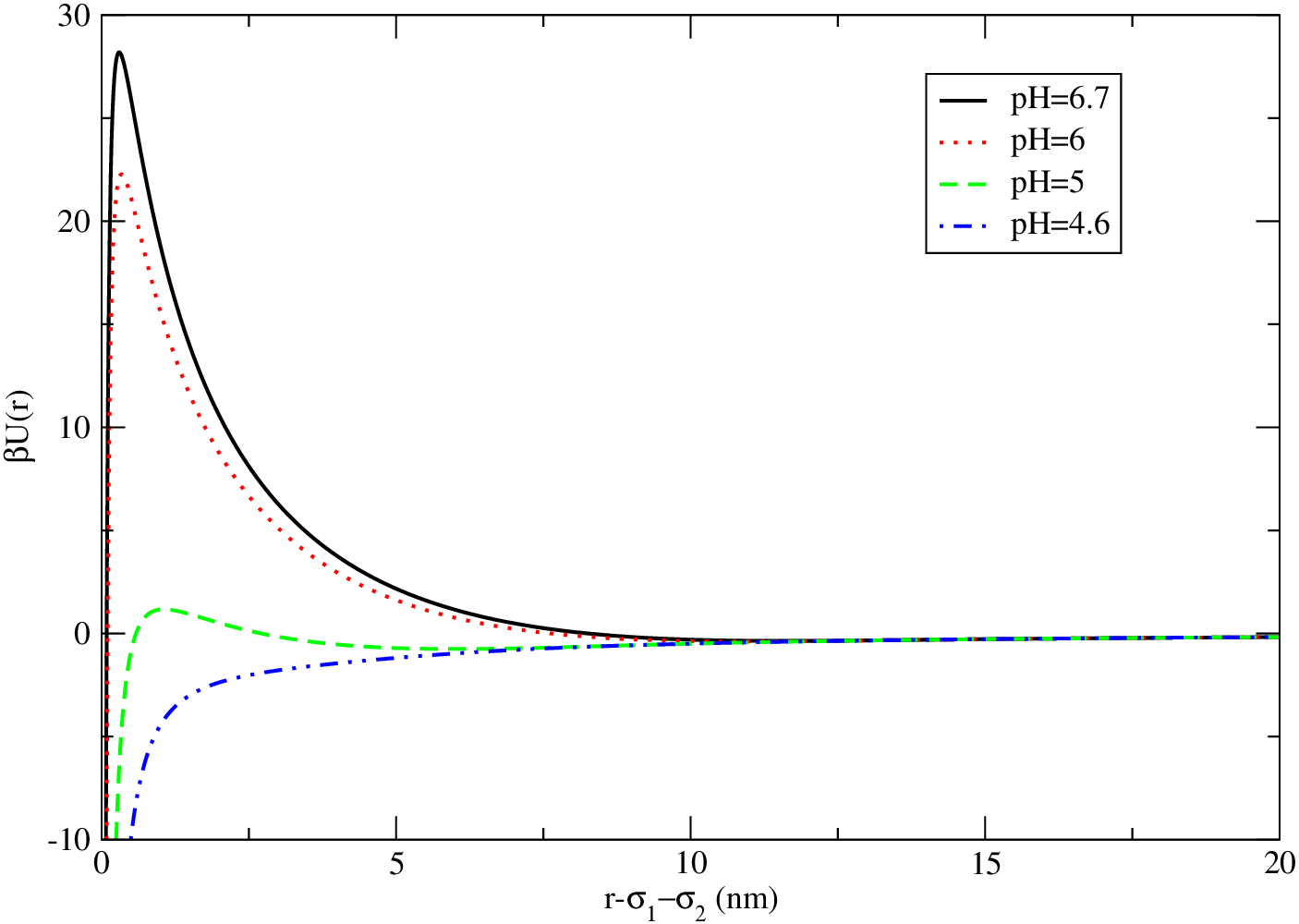}
\caption{Interaction potential between casein micelles with $r=100$ nm for $pH=6.7$ (continuous), 
$6$ (dotted line), $5$ (dashed line) and $4.6$ (dot-dashed line).}
\label{isoel}
\end{figure}

However, this strong repulsion is very sensitive to the value of pH 
because, as pH is decreased, the higher concentration of H ions 
tends to screen negative charges in k-Cn segments, which leads to 
a decrease in the effective length of the polyelectrolyte brush. 
An illustration of the effect of pH in the total interaction potential 
and how the aggregation transition can occur as pH value is approaching 
to the isoelectric point, is presented in figure \ref{isoel} where
the interaction potential is plotted for different values of pH for the same micelles 
considered in figure \ref{pot}. 
The notorious reduction in the steric repulsion gives rise to a net attraction 
at pH lower than 4.6, where van der Waals attraction is responsible for 
irreversible aggregation. 
This behavior suggests that this 
potential could describe correctly the aggregation transition in Montecarlo simulations, 
since the repulsive barrier in the potential energy landscape 
disappears around the pH value known as the isoelectric point of caseins in milk. 

We perform a computer simulation study employing the Montecarlo algorithm 
of Metro-
polis\cite{smit2001understanding} with 500,000 simulation steps with $\Delta t = 0.02$ for a 
system of 1000 particles and a size distribution according to Eq. \ref{distribution}. 
The study consisted in localizing the pH at which aggregation starts by varying
three parameters, specifically, the casein volume fraction, 
the temperature and the grafting density. 
For simplicity, we have not consider the fat globules in the system, 
but only casein-casein interactions.
The set of parameters were chosen in order to explore typical skim milk values.
The reported volume fraction of casein micelles in milk is $=0.13$. 
So, we perform simulations for about a half and the double, specifically, 
$\phi=0.06, 0.13$ and $0.25$, in order to study the effect of the concentration 
of micelles in transition. The temperatures explored in this study are $270 K$, 
$300 K$ and $330 K$ which corresponds to freezing temperature, room temperature, 
and heating of milk.
And finally, the grafting density estimated for CM was $\sigma=0.006$ \cite{de2005stabilization}, 
hence, we studied the aggregation transition at the half and the double,
$\sigma=0.003$ and $0.012$.

\section{Results}
The static structural information of the system, contained in the radial distribution function (RDF), 
$g(r)$, served as the criterion to determine whether the system was in fluid phase or in 
an incipient stage of the formation of a gel.
 
\begin{figure}
\includegraphics [width=8cm]{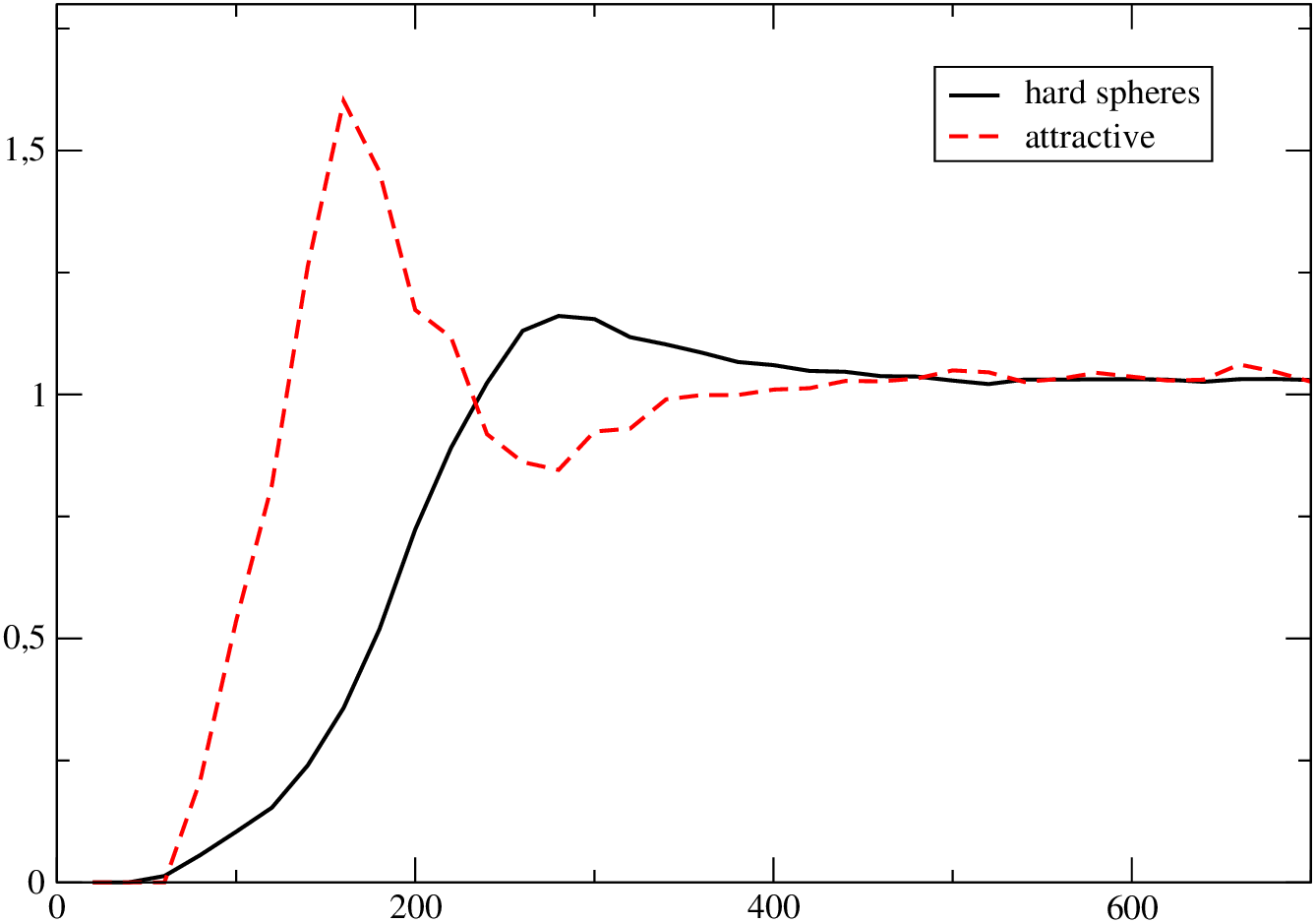}
\caption{Radial distribution function for a polydisperse hard sphere system (continuous line) 
and a purely repulsive system (dashed, red online). 
This static property can serve as a criterion
to locate the aggregation transition in repulsion + attraction systems.}
\label{gdr}
\end{figure}

Figure 3 shows the main differences between fluid and gel phase. 
In order to make more noticeable the main differences exhibited by the RDF between 
fluid and gel phase, in figure \ref{gdr} we have plotted the RDF of a purely attractive 
system where irreversible aggregation occurs (dashed), and a non-interactive system, 
where aggregation will never occur; specifically, a system of hard spheres (continuous); 
both with the same size-distribution. 
As it is well known, the first maximum of the RDF indicates the distance between 
centers at which it is more frequent to find particles, and so, 
for monodisperse hard sphere systems, this equals the diameter of the particles, 
i.e., the center to center distance at contact. 
In this case, since it is not a monodisperse system, 
the most common center to center distance depends on the size distribution, 
but opposite to monodisperse systems, the maximum is much wider. 
The width of this maximum indicates the contribution 
of all distances at which micelles of different sizes can be found. 
After that, the curve does not show typical oscillations as it 
occurs in monodisperse systems, instead, it shows a monotonic decrease.
This behavior in the static structure parameters is well known for polydisperse systems \cite{frenkel1986structure}.
On the opposite side, dashed curve has a very notorious 
maximum located close to the average diameter, which is a sign of aggregation, because 
many particles are in contact with others, and after that, there is a minimum which 
indicates that there is a vacuum of particles in the space surrounding every cluster formed by
several particles. 
In summary, an unambiguous sign of aggregation is the transition 
from a g(r) with just one maximum at distances larger than the average size, 
to the appearance of a pronounced maximum at shorter distances.

\begin{figure}
\includegraphics [width=8cm]{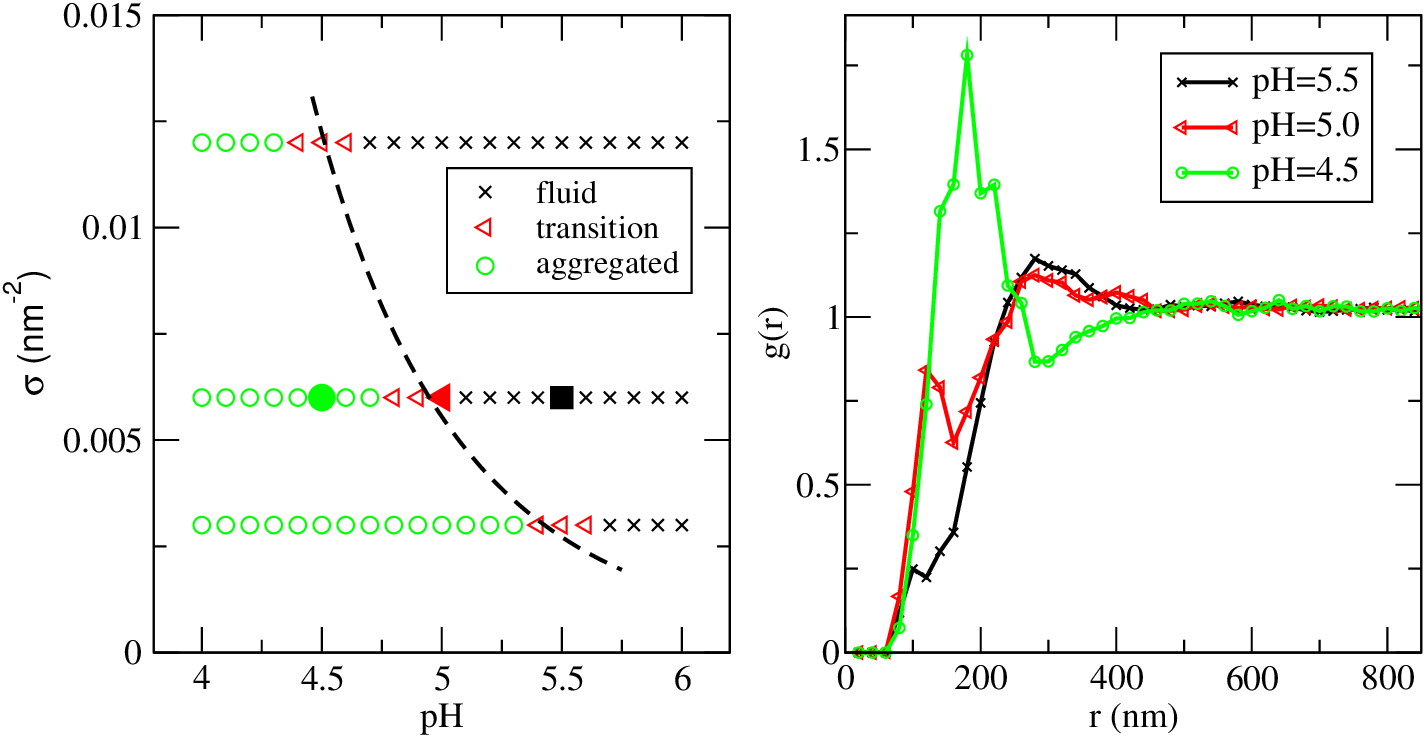}
\caption{Effect of the grafting density $\sigma$ in the aggregation transition. 
At the left panel it is shown the aggregation diagram. 
For different values of $\sigma$ it can be localized the pH at which transition from
disperse (crosses) to aggregated systems (circles) occurs. 
Triangles represent conditions at which aggregation is incipient.
At the right panel, with the same symbol (color online) code, there are shown the RDF of the systems
represented as filled symbols in the transition diagram.}
\label{sigma}
\end{figure}

\begin{figure}
\includegraphics [width=8cm]{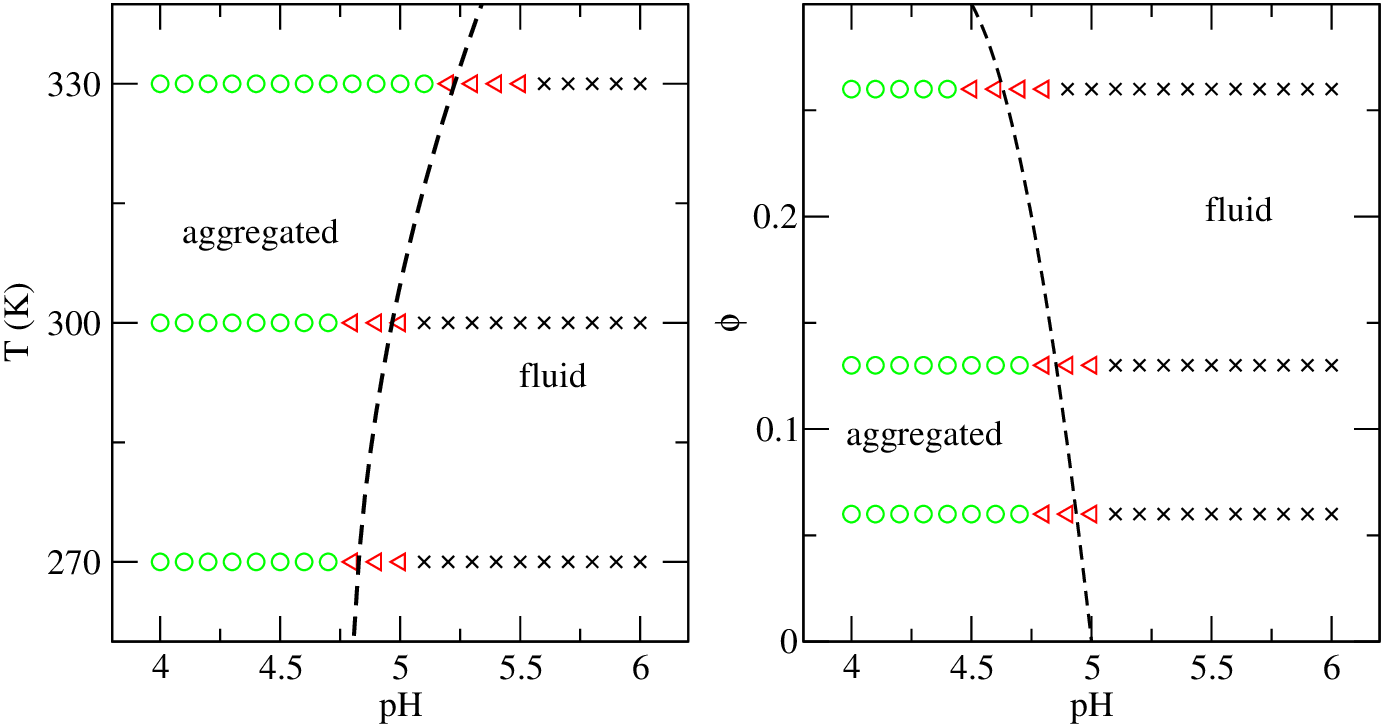}
\caption{Effect of temperature $T$ (left diagram) and volume fraction (right pannel)
on the aggregation transition. The symbols code is the same as in Fig.  \ref{sigma}.}
\label{temperature} 
\end{figure}

Employing the RDF as the criterion to locate the gelification transition, 
we plotted the aggregation diagrams for different conditions in figures \ref{sigma} and \ref{temperature}. 
In this diagrams, crosses represent systems in which the micelles are dispersed in a fluid phase,
while circles those in which clusters are predominant. 
Using the RDF as indicator of aggregation, allows us to locate 
those conditions at which aggregation is incipient, these points are represented as triangles.

Figure \ref{sigma} shows the dependence of the transition on the grafting density 
of the exposed segments of $\kappa$-Cn, $\sigma$, at room temperature, $T=300 K$, and volume 
fraction similar to that of caseins in milk, $\phi=0.13$. In the right panel, 
we show the RDF for three systems (filled symbols)  with $\sigma=0.006$ for different 
pH values, 5.5, 5 and 4.5. 
Observation of these RDF shows how a 
reduction in pH leads the system from a situation (pH=5.5) in which just a few of the smallest micelles 
are aggregated (a short peak around $r=100 nm$) and the overall static structure is practically the 
same as that in hard sphere systems, to a situation (pH=4.5) where there are many clusters 
formed (sharpen peak close to $r=200$); compare this RDFs to those in Fig. \ref{gdr}. 
By direct observation of these 
RDF we sketch the transition diagram (left pannel) for different grafting density in order 
to make explicit the dependence of the stabilizing steric repulsion on the number of 
exposed segments of $\kappa$-Cn.  
As one would expect, it is needed to acidify more the system 
in order to make micelles aggregate for higher values of $\sigma$, because it requires 
more positive ions to neutralize the negative charge of $\kappa$-Cn in order to make the 
polyelectrolyte brush more flexible and penetrable. Triangles in the diagram represent 
those systems in which the number of particles aggregated in clusters has increased, 
but the overall structure still remains being one of a hard sphere system, 
that is why we have labeled this kind of systems as closer to the transition line. 
The dashed line in the diagram is just a guide for the eye of what could be the 
gelification transition line. 

In figure \ref{temperature}, it is shown the dependence of the transition on the 
temperature (left) and on the volume fraction (right). 
As it was shown in figure 2, the interaction potential has an energetic barrier 
as a result of the combination of steric + electrostatic repulsion and van der Waals attraction. 
Since the interaction potential is barely attractive for long distances and strongly repulsive 
for short distances, the micelles behave like hard spheres for higher values of pH but, 
as the positive ions start to neutralize repulsion, 
the intensity of the potential barrier decreases and it is possible for micelles to 
aggregate. 
Another possibility, nevertheless, is that micelles have enough kinetic energy to
overcome the barrier.
In other words, as temperature increases, 
it is more probably that two micelles aggregate after a collision, which would make
the transition easier to occur.
In the left pannel of Fig. \ref{temperature}, we can see that for higher temperatures, 
the transition can occur at higher pH. 
This diagram corresponds to $\phi=0.13$ and $\sigma=.006$.

The effect of micellar concentration on the transition, at constant temperature and $\sigma$ 
($T=300 K$ and $\sigma=.006$), is shown in the right pannel of this figure. 
One could expect that, increasing the collision 
frequency (bacause micelles are closer to each other), the transition would occur at higher pH values, 
but actually, increasing the number of micelles per volume unit, means a higher concentration of positive 
ions needed to neutralize repulsions, so, aggregation is more difficult 
to occur if we increase the micelles concentration. 
For this reason, an increase in micelles concentration, 
would lead the transition to occur at lower pH values, as it is shown in the aggregation diagram.

\section{CONCLUSIONS}
In this work, it has been implemented the interaction potential 
proposed by Tuinier and Kruif \cite{tuinier2002stability, de2005stabilization}
for casein micelles in a Montecarlo computer 
simulation in order to study the dependence of the aggregation 
transition in the case of acid coagulation on the parameters of the system. 
We proposed the radial distribution function as an indicative of the transition. 
Even though in polydisperse systems it is difficult to extract with some detail 
features of the internal structure of a fluid, it has been shown that a comparison
with the RDF of a hard sphere and a sticky hard sphere systems, respectively, 
can serve as a guide to elucidate information concerning the aggregation state
of the suspension.
  
The aggregation diagrams obtained suggest that the interaction potential employed
describes qualitatively correct the aggregation transition of 
a suspension of micelles, for different conditions, in particular, 
from these simulations, it can be determined the dependence of the 
isoelectric point as a function of temperature, casein micelles 
concentration and the surface density of $\kappa$-casein segments. 
The aggregation transition predicted in simulations occurs at higher pH values 
as temperature increases because casein micelles have more kinetic energy which 
allow them to overcome the energetic barrier of the potential. 
On the other hand, the acid coagulation transition 
seems to be very sensitive to the grafting density of $\kappa$-casein, as can be expected 
since the main contribution to the repulsive part of the potential, and hence, 
to the stability of a casein suspension, is due to the steric repulsion of the $\kappa$-casein 
polyelectrolyte brush in the micelle surface. 
Finally, the dependence of the isoelectric point with the concentration of casein micelles predicted, 
although apparently counterintuitive, is full of sense: the more the volume fraction of micelles, 
the lower the pH value at which casein aggregate, because it is needed more 
quantity of positive ions in order to screen the repulsive interaction of the 
polyelectrolyte brushes.
 
Even though this model succeeds in describing the effect of pH on micellar interactions,
and so, in predicting the initial stages of the aggregation transition due to 
a decrease in free energy, it cannot describe the rheological properties of 
dairy products, nor the time evolution of these properties either. 
For this, it is required the implementation of more elaborate models 
at the nanoscopic domain \cite{horne2006casein}.
Although this model is quite simple and contains several assumptions, 
and some of the parameters have been estimated which is difficult to 
avoid in such a complex system like the casein micelle, the description 
provided by this potential and the simulation methodology proposed here, 
can be helpful for a better understanding of the physicochemical 
phenomena ocurring in dairy products.

ACKNOWLEDGMENTS
 
This work was supported by the Consejo Nacional de 
Ciencia y Tecnolog\'{\i}a through grant CB-2010-C01-156423.

\bibliography{MCarxiv}
\end{document}